\documentclass[12pt,nohyper,notoc]{JHEP3}

\usepackage{amsmath,euscript,array,amssymb,cite} 
\input epsf

\def\d {\partial}

\def\H {\mathcal{H}}

\def\L {\mathcal{L}}\def\N {\mathcal{N}}

\def\Z {\mathcal{Z}}
\def\a {\alpha}
\def\b {\beta}

\def\om {\omega}

\def\De {\Delta}
\def\Ga {\Gamma}

\def\ra {\rightarrow}

\def\BE {\begin{equation}}
\def\EE {\end{equation}}
\def\BEA {\begin{eqnarray}}
\def\EEA {\end{eqnarray}}

\def\ms{{\mathfrak M}}
\def\BI{\boldsymbol{I}}

\newcommand{\EQ}[1]{\begin{equation} #1 \end{equation}}
\newcommand{\AL}[1]{\begin{subequations}\begin{align} #1
\end{align}\end{subequations}}
\newcommand{\SP}[1]{\begin{equation}\begin{split} #1 \end{split}\end{equation}}

\title{A Comment on the $\boldsymbol{\chi}_{\boldsymbol y}$ 
Genus and Supersymmetric Quantum Mechanics}

\author{Timothy~J.~Hollowood and Tom~Kingaby\\
Department of Physics, University of Wales Swansea,
Swansea, SA2 8PP, UK\\
E-mail: {\tt pyth@swan.ac.uk}, {\tt pytk@swan.ac.uk}}

\abstract{In this note we show that a simple modification of
  supersymmetric quantum mechanics involving a
  mass term for half the fermions naturally leads to a derivation of
  the integral formula for the $\chi_y$ genus, a quantity that interpolates
  between the Euler characteristic and arithmetic genus. We note that
  this modification naturally arises in the moduli space dynamics of
  monopoles or instantons in theories with 16 supercharges partially
  broken to 8 supercharges by mass terms.}
\keywords{} \preprint{{\tt hep-th/0303018}\\SWAT-}

\begin{document}

Index theory, as developed by Atiyah and Singer \cite{Atiyah},  
has many important applications in theoretical physics. For the
physicist, many of the
mathematical complications of index theory can be avoided by following
Alvarez-Gaum\'e \cite{AG} (see also \cite{Friedan:1983xr}) 
and formulating it
in the context of a suitable supersymmetric quantum mechanical
system. The crucial idea is to find a supersymmetric quantum
mechanical system whose Witten index yields the topological index of
the elliptic complex in question.

On the other hand, there are situations in which supersymmetric
quantum mechanics arises naturally. The one we have in mind here, is
in the semi-classical quantization of solitons in field theory. In the
classical limit the dynamics can often be described in the terms of
motion on the moduli space of the soliton (the space of classical
solutions). Semi-classical effects are then described by quantum
mechanics on the moduli space. In a supersymmetric theory, soliton
solutions generally preserve half the supersymmetries of the parent
theory and these are inherited by the quantum mechanical system.

An example of this set up is in five-dimensional gauge theories which
have soliton solutions consisting of conventional instanton solutions
embedded in the four spatial dimensions. Semi-classical effects are
described by quantum mechanics on the moduli space of Yang-Mills
instantons of a given charge $\ms$
\cite{review}. If the parent theory has 16 supercharges ($\N=4$ in four
dimensions) then the quantum mechanical system is the one associated
to the de Rahm complex which usually admits 2 supercharges but since
$\ms$ is hyper-K\"ahler this is enhanced to 8. On the other hand if
the parent theory has 8 supercharges ($\N=2$ in four dimensions) then
the quantum mechanical system is the one associated to the Dolbeault---or
equivalently, since $\ms$ is hyper-K\"ahler, the Dirac---complex 
which usually admits 1 supercharge but, as above, this is enhanced to 4.
A similar application concerns the semi-classical
quantization of monopoles in the same theories in four dimensions. In
this case the same kind of quantum mechanical systems arise but now
associated to the monopole moduli space (see, for example,
\cite{Gauntlett:yj,Gauntlett:1993sh,Gauntlett:2000ks}). 

It is well known that the gauge theories with 16 supercharges can
be broken to one with 8 supercharges by adding suitable mass
terms. In four dimensions the mass deformed theory is sometimes called
the $\N=2^*$ theory. These mass terms induce terms in
the quantum mechanical system describing the soliton dynamics which
have the effect of breaking one half of the supersymmetries. This
suggests that the system with the supersymmetry-breaking terms will
somehow interpolate between the de Rahm complex and Dolbeault complex 
and, in particular, its Witten Index will interpolate between the
associated topological indices; that is the {\it Euler characteristic\/} and
the {\it arithmetic genus\/}. In the applications to 
instanton-solitons 
in five-dimensional supersymmetric gauge theory the index is directly
relevant because it determines the instanton contributions to the
prepotential.  

Although the problem
was inspired by instanton solitons in five-dimensional gauge theory
(or monopoles in the associated four-dimensional theory)
we shall divorce our discussion from these particular examples because
there are additional complications in these cases. In particular, the
moduli spaces of instantons or monopoles are 
non-compact and this leads to subtleties
in defining the Witten Index. We shall work with a target space $\ms$
which is compact. In addition, in both the instanton and monopole examples,
VEVs for scalar fields in the parent theory lead to more complicated
quantum mechanical systems involving potentials induced by 
vector fields on the moduli
space, as one can see in the instanton case in \cite{review} and in the
monopole case in \cite{Gauntlett:2000ks}. This leads to an 
equivariant generalization of 
index theory and once again we shall avoid these complications in this
note.

Following Alvarez-Gaum\'e \cite{AG}, 
we start with the quantum mechanical system associated to the de Rahm
complex of a compact manifold $\ms$. Let $X^\mu$, $\mu=1,\ldots,n$, 
be local coordinates for $\ms$ which become
one-dimensional fields $X^\mu(t)$ in the quantum mechanical system. 
Associated to these bosonic quantities, we have 2-component 
fermions $\psi^\mu_\alpha(t)$, $\alpha=1,2$,
which are Grassmann-valued fields. The basic Lagrangian that defines
the system is 
\EQ{\L=\tfrac12 g_{\mu\nu}\dot{X}^\mu\dot{X}^\nu+
    \tfrac{i}{2}g_{\mu\nu}\psi^\mu_\a \dot{\psi}^\nu_\a+
    \tfrac{i}{2}g_{\mu\nu}\psi^\mu_\a \Ga^\nu_{\sigma\rho}
\dot{X}^\sigma\psi^\rho_\a+
    \tfrac{1}{4}R_{\mu\nu\sigma\rho}\psi^\mu_1\psi^\nu_1
\psi^\sigma_2\psi^\rho_2\ ,
\label{GenL}
}
where $g_{\mu\nu}(X)$ is the metric on $\ms$ and
$R_{\mu\nu\sigma\rho}(X)$ is the usual Riemann tensor associated to
the Levi-Civita connection $\Gamma_{\sigma\rho}^\nu(X)$. 

The quantization of the theory follows by imposing the following canonical
(anti-)commutation relations:
\EQ{
\big[X^\mu,p^\nu\big]=ig^{\mu\nu}\ ,\qquad
\big\{\psi^\mu_a,\psi_b^\nu\big\}=\delta_{ab}g^{\mu\nu}\ .
}
It is useful to define the super-covariant momentum 
\EQ{
\pi_\mu=p_\mu+\Gamma_{\mu\nu\lambda}\psi^\nu_1\psi^\lambda_2\ .
}
The system is invariant under 2 supersymmetries generated by the supersymmetry
charges 
\AL{
Q_\alpha=\psi^\mu_\alpha\pi_\mu\ .
}
It is important to realize that operator ordering here is
significant. The Hamiltonian is obtained by the anti-computation of
two super-charges:
\EQ{
\big\{Q_\alpha,Q_\beta\big\}=2\delta_{\alpha\beta}\H\ ,
}
yielding
\EQ{
\H=\frac1{2\sqrt g}\pi_\mu\sqrt gg^{\mu\nu}\pi_\nu\ .
}

The Hilbert space of the model is realized in terms of a fermionic
Fock space, with creation operators and annihilation operators given
by the combinations
\EQ{
b^{\mu\dagger}=\tfrac1{\sqrt2}(\psi^\mu_1-i\psi^\mu_2)\ ,\qquad
b^{\mu}=\tfrac1{\sqrt2}(\psi^\mu_1+i\psi^\mu_2)\ .
}
The states
\EQ{
f_{\mu_1\cdots\mu_p}(X)b^{\mu_1\dagger}\cdots b^{\mu_p\dagger}|0\rangle
}
are in one-to-one correspondence with the de Rahm complex of $\ms$;
for instance, the above state corresponds to the $p$-form
\EQ{
f_{\mu_1\cdots\mu_p}(X)\,dX^{\mu_1}\wedge\cdots\wedge dX^{\mu_p}\ .
}
Under this correspondence the supercharges $Q_\alpha$ are realized in
terms of the exterior derivative $d$ and its adjoint $d^\dagger$:
\EQ{
Q_1=-\tfrac i{\sqrt2}(d-d^\dagger)\ ,\qquad
Q_2=\tfrac1{\sqrt2}(d+d^\dagger)\ .
}

We now suppose that $\ms$ is a K\"ahler manifold for which $g$ is the
K\"ahler metric. So $\ms$ has a K\"ahler
form $\omega$ that is closed. The K\"ahler metric $g$ is Hermitian with
respect to the complex structure $\BI$ and furthermore 
$g(\BI X, Y)=\omega(X,Y)$, for 2 arbitrary tangent vectors $X$ and $Y$. 
Under these circumstances, it is well known that the quantum
mechanical system admits 2 additional supersymmetries generated by the
supercharges
\EQ{
Q'_\alpha=(\BI\cdot\psi_\alpha)^\mu\pi_\mu\ .
}
Since $\ms$ is a complex manifold we can choose (anti-)holomorphic
coordinates $(z^j,\bar z^j)$, $j=1,\ldots,\tfrac12n$, compatible with
the complex structure: $(\BI\cdot z)^j=iz^j$ and $(\BI\cdot\bar z)^j=
-i\bar z^j$. In this coordinate system, the Hilbert space can be built
on a fermionic Fock space for which 
$\psi_1^j$ and $\bar\psi_2^j$ are the creation operators while
$\bar\psi_1^j$ and $\psi_2^j$ are the annihilation operators. States
in the Hilbert space are naturally identified with elements of the
Dolbeault complex via the correspondence 
\SP{
&\psi_1^{j_1}\cdots\psi_1^{j_p}\bar\psi_2^{k_1}\cdots\bar\psi_2^{k_q}|0\rangle
\longleftrightarrow 
dz^{j_1}\wedge\cdots\wedge dz^{j_p}\wedge d\bar z^{k_1}\wedge\cdots
d\bar z^{k_q}\ .}

Since $\ms$ is K\"ahler, we can add a kind of mass for one of the
species of fermions to the Lagrangian:
\EQ{
\L_m=-\tfrac12 m
\omega(\psi_2,\psi_2)+c=-\tfrac12m\psi^\mu_2\omega_{\mu\nu}\psi_2^\nu+c
=-\tfrac 12m\psi_{2\mu}(\BI\cdot\psi_2)^\mu+c\ ,
\label{pert}
}
where $m$ is a parameter. In the application to instantons, such a
term was derived in the effective quantum mechanics on the moduli
space in \cite{review}. In the monopole application, such a term can be
extracted indirectly from \cite{Gauntlett:2000ks}, by choosing the
matter content of the $\N=2$ theory to transform in the adjoint
representation. In both cases $m$ is the mass of the adjoint
hypermultiplet in the parent theory.

The term \eqref{pert} is only invariant under half the original 
supersymmetries; namely those generated by $Q_1$ and $Q'_1$, 
following from the fact that the K\"ahler form is
covariantly constant on a K\"ahler manifold. In \eqref{pert}, $c$ is a
constant that arises via a normal ordering prescription and its value 
is fixed as follows. In the canonical formalism, we require that
the term \eqref{pert} leads to a modification of the Hamiltonian operator of
the normal-ordered form
\EQ{
\H'=\H+\H_m\ ,\qquad\H_m=\tfrac12 m:\psi_{2\mu}(\BI\cdot\psi_2)^\mu:\ ,
\label{perth}
}
in order that it annihilates the vacuum state $|0\rangle$. 
Notice in the language of the Dolbeault complex, $\H_m$, up
to the factor of $m$, simply counts the anti-holomorphic
degree. Ensuring that \eqref{pert} leads to \eqref{perth}, fixes
\EQ{
c=mn/2\ .
}
With the mass term added, the supersymmetry algebra gains a central
charge:
\EQ{
Q_1^2=\H'-\Z\ ,\qquad Q_1^{\prime2}=\H'-\Z\ ,\qquad\big\{Q_1,Q'_1\big\}=0\ ,
}
where it is immediately apparent---since $Q_1$ is unchanged---that 
\EQ{
\Z=\H_m\ .
}

The question is what does the modification do to the Witten Index of the
model? Since we have remarked that $\H_m$
has a very simple action on the Dolbeault
complex---it simply counts the anti-holomorphic degree of a
form multiplied by $m$---the Witten Index of the deformed
system will be given by
\EQ{
{\rm ind}_W=\sum_{i,j}(-1)^{i+j}b_{i,j}e^{-\beta m j}\ ,
\label{witi}
}
where $b_{i,j}$ are the Betti numbers of the Dolbeault complex with $i$
and $j$ being the holomorphic and anti-holomorphic degrees,
respectively.

So the addition of the term \eqref{pert} or \eqref{perth}, 
which breaks half the
supersymmetry, is to deform the index. When $m=0$ we recover the Euler
characteristic. In the limit, $m\to\infty$, the index reduces to the
arithmetic or Todd genus $\sum_{i}(-1)^ib_{i,0}$. In fact the interpolating
quantity \eqref{witi} is the $\chi_y$-genus of Hirzebruch \cite{Hirz}
\EQ{
\chi_y=\sum_{i,j}(-1)^iy^jb_{i,j}\ ,\qquad y=-e^{-\beta m}\ .
}
Now we see how it can naturally be obtained from a deformed quantum
mechanical supersymmetric $\sigma$-model. Note that the $\chi_y$-genus
has also been related to supersymmetric quantum mechanics in \cite{meng},
via a twisted of the boundary conditions on the fermions. Our
approach obtained by adding the deformation \eqref{pert}, motivated by 
the application to soliton dynamics, is different. In related work,
a geometric interpretation of the $\chi_y$-genus has
been given for hyper-K\"ahler geometries in
\cite{Thompson:1999iw,Thompson:1998vx}. In particular, this is
relevant for the application of our results to soliton quantization
for which the relevant geometry is hyper-K\"ahler. 

It remains to derive the known 
integral expression for the $\chi_y$-genus by
computing the partition function of the deformed quantum mechanical
system. The steps are a simple generalization of the standard
derivation of index densities of supersymmetric quantum mechanics \cite{AG}.
As usual the Witten Index can be calculated by computing the Euclidean
functional integral with fields being periodic in $t$. We define
$\beta$ to be the period. The resulting functional integral expression is then 
independent of $\beta$---apart from via the
combination $\beta m$---and may be readily evaluated in the limit
$\beta\to0$ (with fixed $\beta m$). 
In this limit, constant configurations of $X^\mu(t)$ and
$\psi_\alpha^\mu(t)$ 
dominate the functional integral and one can integrate out the
fluctuations to Gaussian order. 
To this end we expand around constant configurations: 
\EQ{ 
X^\mu\ra
x^\mu+\delta X^\mu(t)\ ,\qquad
\psi^\mu_\a\ra\eta^\mu_\a+\delta\psi^\mu_\a(t)\ .
}
We can now 
integrate out the fluctuations separately and this is greatly
facilitated by choosing at each 
$x^\mu$ normal co-ordinates for which:
\EQ{
g_{\mu\nu}(x)=\delta_{\mu\nu}+{\cal O}(\delta X^2)\ .
\label{nc}
}
The Euclidean action then splits in two:
\EQ{
S=S_c+\int_0^\b dt\, \L_f\ .
}
The constant part is
\EQ{
S_c=-\tfrac14\beta
R_{\mu\nu\sigma\rho}\eta^\mu_1\eta^\nu_1\eta^\sigma_2\eta^\rho_2
+\tfrac12\beta m\eta^\mu_2\om_{\mu\nu}\eta^\nu_2+\tfrac12\beta nm\ ,
}
where the final term arises from the normal-ordering constant in \eqref{pert}.
This expression implies that the fermions zero-modes $\eta_1^\mu$ scale
like $\beta^{-1/2}$, while $\eta_2^\mu$ do not scale with $\beta$
(remember that $\beta m$ is fixed). The fluctuation part is then 
\EQ{
\L_f=\tfrac12 \delta
X^\mu \De^B_{\mu\nu}\delta X^\nu
+\tfrac{1}{2}\delta\psi^\mu\Delta^F_{\mu\nu}\delta\psi^\nu+\cdots\ ,
}
where the ellipsis represent non-Gaussian terms which only contribute
at a higher order in $\beta$ and hence can be ignored. Using the fact
that in normal coordinates \eqref{nc}
\EQ{
\psi^\mu_1\Gamma_{\mu\sigma\rho}(X)\dot X^\sigma\psi^\rho_1
=\tfrac12\delta
X^\mu R_{\mu\nu\sigma\rho}(x)\eta^\sigma_1\eta^\rho_1\delta\dot X^\nu+\cdots
}
to leading order and up to total derivatives, the bosonic operator is
\EQ{
\De^B_{\mu\nu}=-\delta_{\mu\nu}\d_t^2-
\tfrac{1}{2}R_{\mu\nu\sigma\rho}\eta^\sigma_1\eta^\rho_1\d_t+\cdots\ ,
} 
to leading order in $\beta$. The fermionic operator is matrix-valued:
\EQ{ 
\De^F_{\mu\nu}=\left(\begin{array}{cc}
\delta_{\mu\nu}\d_t
& 0\\ 0 &
\delta_{\mu\nu}\d_t+\frac{1}{2}R_{\mu\nu\sigma\rho}\eta^\sigma_1\eta^\rho_1
-\beta m\om_{\mu\nu}
\end{array}\right)+\cdots\ ,
}
to leading order in $\beta$.

We can now integrate out the fluctuations $\delta X^\mu$ and $\delta\psi^\mu$,
as well as the constant modes $\eta_2^\mu$ keeping careful track of
the overall normalization of the functional integral. As usual we can
write the resulting integral over $X^\mu$ and $\eta_1^\mu$ as an
integral over differential forms by identifying
$\eta_1^\mu=dX^\mu$. Finally we have
\EQ{
{\rm ind}_W=
\Big(\frac i{2\pi}\Big)^{n/2}e^{-n\beta m/2}\int_\ms
    {\rm det}^{1/2}\left(\frac{R/2
\sinh(R/2-\beta m\om/2 )}
{\sinh(R/2)}\right)\ ,
}
where $R$ is the matrix-valued curvature 2-form. This can be written as 
\EQ{
{\rm
  ind}_W=\int_{\ms}\prod_{i=1}^{n/2}\frac{x_i(1+ye^{-x_i})}{1-e^{-x_i}}\ ,
}
where $x_i$ are the skew eigenvalues of $R/4\pi$ and
$y=-e^{-\beta m}$. This reproduces the integral form for the $\chi_y$
genus \cite{Hirz}.


\begin{thebibliography}{99}

\small{

\bibitem{Atiyah}
M.~F.~Atiyah and I.~M.~Singer,
Annals Math.\  {\bf 87} (1968) 484;
Annals Math.\  {\bf 87} (1968) 546;
Annals Math.\  {\bf 93} (1971) 119;
Annals Math.\  {\bf 93} (1971) 139.




\bibitem{Hirz}
F.~Hirzebruch, ``{\sl Topological Methods in Algebraic Geometry\/},''
3rd Edition, Springer-Verlag, Berlin 1966.\\
F.~Hirzebruch and D.~Zagier, ``{\sl The Atiyah-Singer Theorem and
Elementary Number Theory\/},'' Mathematical Lecture Series 3, Publish
or Perish, Boston 1974.


\bibitem{AG}
L.~Alvarez-Gaume,
Commun.\ Math.\ Phys.\  {\bf 90} (1983) 161;
J.\ Phys.\ A {\bf 16} (1983) 4177.

\bibitem{Friedan:1983xr}
D.~Friedan and P.~Windey,
Nucl.\ Phys.\ B {\bf 235} (1984) 395.


\bibitem{review}
N.~Dorey, T.~J.~Hollowood, V.~V.~Khoze and M.~P.~Mattis,
Phys.\ Rept.\  {\bf 371} (2002) 231
{\tt[arXiv:hep-th/0206063]}.

\bibitem{Gauntlett:yj}
J.~P.~Gauntlett,
Nucl.\ Phys.\ B {\bf 400} (1993) 103
[arXiv:hep-th/9205008].

\bibitem{Gauntlett:1993sh}
J.~P.~Gauntlett,
Nucl.\ Phys.\ B {\bf 411} (1994) 443
[arXiv:hep-th/9305068].



\bibitem{Gauntlett:2000ks}
J.~P.~Gauntlett, C.~j.~Kim, K.~M.~Lee and P.~Yi,
Phys.\ Rev.\ D {\bf 63} (2001) 065020
{\tt[arXiv:hep-th/0008031]}.

\bibitem{meng}
G.~Meng, 
J.\ Phys.\ A {\bf34} (2003) 1083.

\bibitem{Thompson:1999iw}
G.~Thompson,
Commun.\ Math.\ Phys.\  {\bf 212} (2000) 649.

\bibitem{Thompson:1998vx}
G.~Thompson,
Adv.\ Theor.\ Math.\ Phys.\  {\bf 3} (1999) 249
[arXiv:hep-th/9811199].



}\end{thebibliography}
\end{document}